\documentclass[12pt]{article}

\title{\bf 
Properties of Higher-Order \\Phase Transitions }
\author{ 
{\it W. Janke}\\
Institut f\"ur Theoretische Physik,\\
Universit\"at Leipzig,\\
Augustusplatz 10/11, \\
04109 Leipzig, Germany \\
{}\\
{\it D.A. Johnston}\\ 
Department of Mathematics,\\
School of Mathematical and Computer Sciences,\\
Heriot-Watt University,\\
Riccarton,\\
Edinburgh, EH14 4AS, Scotland \\
{}\\and \\
{}\\
{\em R. Kenna}\\
Applied Mathematics Research Centre,\\
Coventry University,\\
Coventry, CV1 5FB, England
}
\begin{document}
\maketitle
                      {\Large
                      \begin{abstract}
%
Experimental evidence for the existence of strictly higher-order phase transitions 
(of order three or above in the Ehrenfest sense) is tenuous at best.
However, there is no known physical reason why such transitions should not exist in nature. 
Here, higher-order transitions characterized by both discontinuities and divergences 
are analysed through the medium of partition function zeros. 
Properties of the distributions of zeros are derived,  certain scaling relations 
are recovered, and new ones are presented.

%
                        \end{abstract} }
%
  \thispagestyle{empty}
%
%
  \newpage
%
                  \pagenumbering{arabic}

\section{Introduction}
\label{intro}
\setcounter{equation}{0}

In its original format, 
the Ehrenfest classification scheme identifies the order of a phase transition as
that of the lowest derivative of the Helmholtz free energy which displays a 
discontinuity there \cite{Eh33}. Typical transitions which fit to this scheme are 
first-order solid-liquid-vapour transitions and second-order superconducting 
transitions.
There are, however, many transitions characterised by divergent 
 rather than  discontinuous behaviour. 
Examples include ferromagnetic transitions in metals and the spontaneous
symmetry breaking of the Higgs field in particle physics, which display 
power-law or logarithmic divergent behavior as the transition is approached.
The classification scheme has, in practice, been extended to encompass these
scenarios and the order of a transition is commonly given by the 
order of the lowest derivative in which any type of non-analytic behaviour
is manifest.

It has long been suspected that transitions of Ehrenfest order greater than two
(with a discontinuity at the transition point) do not exist in nature. 
However there is no obvious physical reason why this should be the case. 
In fact,  recent experimental observations
 of the magnetic properties of a
cubic superconductor have been ascribed to its
possessing a fourth-order discontinuous transition
\cite{KuDo99}
(see also \cite{WoWr99} where the existence of
well defined anomalies in the specific heat at the transition point was
claimed). A theory for higher-order transitions was developed in 
\cite{Ku97,KuSa02,FaYu04}
and found to be consistent with
experimental work.

Higher-order phase transitions (with either a discontinuity or a divergence 
in an appropriate free-energy derivative) certainly exist in a number of theoretical models.
There are third-order temperature-driven transitions in
various ferromagnetic and antiferromagnetic spin models \cite{BeKa52,spin},
as well as spin models coupled to quantum gravity \cite{Ka86,fat}.
Recent theoretical studies also indicate the presence of third-order transitions in 
various superconductors \cite{CrNo01}, DNA under mechanical strain  \cite{RuBr02},
spin glasses \cite{CrRi03},
lattice and continuum gauge theories \cite{QCD} and matrix models linked to supersymmetry \cite{FuMi04}.
A fourth-order transition in a model of a  branched polymer was studied in \cite{BiBu96}
and the Berezinskii-Kosterlitz-Thouless transition is of infinite-order \cite{KT}.

In this paper, we analyse higher-order transitions through the medium of 
partition function zeros. 
To set the notation, let $t$ represent a generic reduced even variable and $h$ be the
odd equivalent so that $t=T/T_c-1$ and $h=H/k_BT$ in the notation of the Ising model
(i.e., $T$ is the temperature, which is critical at $T_c$, and $H$ is 
the external magnetic field). The critical point is given by $(t,h)=(0,0)$.
This may be the end-point of a line of first-order transitions, as is the case in the 
Ising or Potts models. In the Potts-like case where the locus of transitions 
is curved, we may instead assume that $t$ and $h$ are suitable mixed variables, so that
$h$ is orthogonal to $t$, which parameterizes arc length along the transition line \cite{KaSt00}.
The free energy in the thermodynamic limit is denoted by $f(t,h)$
and its $n^{\rm{th}}$-order even and odd derivatives are
$f^{(n)}_t(t,h)$ and $f^{(n)}_h(t,h)$ so that
the internal energy, specific heat, magnetization and susceptibility are
given (up to some inert  factors) as
$
 e(t,h) =  f^{(1)}_t (t,h)
$,
$
 C(t,h) =  f^{(2)}_t (t,h)
$,
$
 m(t,h) =  f^{(1)}_h (t,h)
$,
and
$
 \chi (t,h) =  f^{(2)}_h (t,h)
$,
respectively. 
In the following, to simplify the notation, 
we drop the explicit functional dependency on a variable
if it vanishes.

One then commonly describes as an $m^{\rm{th}}$-order phase transition a situation where the 
first $(m-1)$ derivatives of the free energy with respect to the even (thermal) 
variable are continuous, but where
the $m^{\rm{th}}$ thermal derivative is singular, with a discontinuity or a divergence
at the transition point. The lowest $(m^\prime-1)$ derivatives of the free energy 
with respect to the odd (field) variable may also be continuous in $t$, 
with a singularity occuring in the ${m^\prime}^{\rm{th}}$ derivative.
Thus a continuous specific heat is realized if $m>2$ and the susceptibility
is also continuous if $m^\prime >2$ as well.
This situation, which is not 
normally possible in a ferromagnet\footnote{
One can readily see this by considering 
the Rushbrooke scaling law (\ref{RG2}) together with 
hyperscaling which give  $\gamma/\nu=d-2 \beta/\nu$ ($d$ being dimensionality and $\nu$ 
the correlation-length critical exponent).
Since for a system of finite linear extent $L$, the magnetisation obeys
$\langle |m| \rangle  \propto L^{-\beta/\nu}$, and since 
completely uncorrelated ferromagnetic spins would lead by the central limit theorem 
to $\langle |m| \rangle  \propto L^{-d/2}$, we obtain the bound $\beta/\nu < d/2$, 
since the actual decay in the correlated case is slower. From this, one obtains the restriction that
$ \gamma/ \nu$ cannot be negative for a ferromagnet.
}, is the one analysed in \cite{Ku97,KuSa02}, in which $m=m^\prime >2$.
Such higher-order transitions may be possible in branched polymers and 
diamagnets such as superconductors.

In the  more common scenario where $m^\prime$ is not necessarily the same as $m$, 
the scaling behaviour at the transition may be 
described by  critical exponents
at $h=0$:
\begin{equation}
 f^{(m)}_t(t) \sim t^{-A} \;, \quad
 f^{(m^\prime)}_h(t) \sim t^{-G} \;, \quad
 f^{(1)}_h(t) \sim t^{\beta} \;,
 \label{3}
\end{equation}
while,  for the magnetization in field at $t=0$, we write 
\begin{equation}
 f^{(1)}_h(h) \sim h^{1/\delta} \;.
\label{4}
\end{equation}
In the familiar case of a second-order transition ($m=m^\prime=2$), the exponents 
$A$ and $G$ become, in standard notation, 
$\alpha$ and $\gamma$, associated with specific heat and 
susceptibility, respectively.

In the theoretical work of \cite{KuSa02}, the 
following scaling relations were derived for the case of a diverging higher-order transition
in which $m^\prime = m$:
\begin{eqnarray}
 (m-1) A + m \beta + G = m(m-1) \;,
\label{Rushbrooke1}
\\
G =  \beta \left( (m-1) \delta -1\right) \;.
\label{Griffiths1}
\end{eqnarray}
In the second-order case  (\ref{Rushbrooke1}) and (\ref{Griffiths1}) become 
equivalent to the standard Rushbrooke and Griffiths scaling laws,
\begin{equation}
 \alpha + 2 \beta + \gamma = 2 \;, \quad
\gamma =  \beta (\delta -1) \;,
\label{RG2}
\end{equation}
as one would expect.

Since the seminal work by Lee and Yang \cite{LY} as well as by Fisher \cite{Fi64}, 
the analysis of zeros of the partition function has become fundamental 
to the study of phase transitions. 
Fisher zeros in the complex temperature plane 
pinch the real axis at the physical transition point.
The locus of Lee-Yang zeros, in the complex magnetic-field plane, 
is controlled by the (real) temperature parameter
and in the high-temperature
phase, where  there is no transition, it ends at the Yang-Lee edge.
Denoting the distance
of the edge from the real axis by $r_{\rm{YL}}$, one has the generic behaviour 
\begin{equation}
 r_{\rm{YL}}(t) \sim t ^{\Delta/2}
 \;,
\label{edge}
\end{equation}
at a second-order transition. The exponent $\Delta$ is related to the other 
exponents through $\Delta = {2 \gamma \delta}/{(\delta -1)}$.

In the remainder of this paper, 
a number of results concerning the locus and density of
zeros are presented.
Higher-order transitions controlled by a single parameter are analysed 
in Sec.~\ref{sec:fisher} where the locii and densities of the corresponding
Fisher zeros are determined. Various restrictions on the properties
of such transitions are established and simple quantitative methods for
analysing them are suggested.
In Sec.~\ref{sec:LY}, the focus is on the zeros of the Lee-Yang variety
where even and odd control parameters come into play.
Here, the scaling relations  (\ref{Rushbrooke1})
and (\ref{Griffiths1}) are recovered and elucidated and a number of other
ones are presented.
Finally, conclusions are drawn in Sec.~\ref{sec:ccl}.

\section{Fisher Zeros}
\label{sec:fisher}
\setcounter{equation}{0}

In the bulk of physical models 
the locus of Fisher zeros is linear 
in a suitable parameter, $u$, which is a function of $t$ and
can be parameterized near the transition point, $u_c$, 
by  \cite{AbeLY,SuzukiLY,AbeF,SuzukiF}
\begin{equation}
u(r) =u_c+ r \exp(i \phi(r))
\; .
\label{singline}
\end{equation}
This singular line  in the upper half-plane has $0 < \phi(r) < \pi$, while that
in the lower half is its complex conjugate.

In the thermodynamic limit,  the (reduced) free energy  is 
\begin{equation}
 f(t) = 2 {\rm{Re}} \int_0^R{\ln{\left( u-u(r) \right)}} g(r) dr \; ,
\end{equation}
where $g(r)$ is the density of zeros and $R$ is a cutoff. 
We are interested in the moments given by
\begin{equation}
 f_t^{(n)}(t) 
 =
 2 (-1)^{n-1}(n-1)!
   {\rm{Re}} \int_0^R{\frac{g(r)}{\left(u-u(r)\right)^n}}  dr 
\; ,
\label{2}
\end{equation}
and consider the cases of discontinuous and divergent $m^{\rm{th}}$-order
temperature-driven transitions separately.

The difference in free energies on either side of 
the transition can be expanded as 
$
 f_+(t) - f_-(t) 
 =
 \sum_{n=1}^\infty{c_n}(u-u_c)^n
$,
where $+$ and $-$ refer to above and below $u_c$. 
For a discontinuous transition, $c_n =0$ for
$n<m$, while $c_m \ne 0$ and the discontinuity in 
the $m^{\rm{th}}$ derivative of the free energy is
\begin{equation}
\Delta f_t^{(m)} = m!c_m
\; .
\label{fm}
\end{equation}
Now,  the real parts of the free energies must match across the singular line
(otherwise the transition would be of order zero)
which, from (\ref{singline}), means 
$
 \sum_{n=m}^\infty{c_n}r^n \cos{n\phi (r)}
 = 0
$.
Therefore 
 the impact angle (in the upper half-plane), $\phi = \lim_{r \rightarrow 0}{\phi(r)}$,
is 
\begin{equation}
 \phi = \frac{(2l+1)\pi}{2m}
\quad \quad {\rm{for}} \quad l = 0,\dots, m-1
\; .
\label{general}
\end{equation}
It is now clear that, under these conditions, 
vertical impact is allowed only at any discontinuous
transitions of odd order.
A discontinuous second-order transition
with impact angle $\pi/2$ is forbidden.
Similarly an impact angle of $\pi/6$, for example, is only allowed
at a transition of order $3$ or $9$ or $15$, etc. 
This recovers disparate results for first-, second- and third-order
transitions in \cite{LY}, \cite{BlEv} and \cite{fat} which are associated 
with  impact angles $\pi/2$ (corresponding to $l=0$),  $\pi/4$ ($l=0$)
and $\pi/2$ ($l=1$), respectively. 
The question now arises as to the mechanism by which the system selects its
$l$-value. One expects that further studies of higher-order transitions will be required to 
provide an answer.

Let
$
 t = u - u_c$, $\tau = t e^{-i\phi}
$
and assume that the leading behaviour of the density of zeros is
$
 g(r) = g_0 r^{p}  
$,
where $g_0 $ is constant.
If $p$ is an integer, analytical extension of 
the integration (\ref{2}) to the complex plane yields the following result for the $n^{\rm{th}}$ derivative:
\begin{equation}
 f_t^{(n)} (t) 
 =
\left.{
 - 2 (n-1)! g_0  {\rm{Re}}
 \sum_{j=1}^{p+1} e^{-in\phi} T_j
}\right|_\delta^R 
\; ,
\label{39}
\end{equation} 
where $\delta$ is a lower integral cutoff
and
\begin{eqnarray}
 T_j & = & \frac{ p!\tau^{p+1-j}(r-\tau)^{j-n}}{(j-1)!(p+1-j)!(j-n)}
\quad \mbox{
for $j \ne n$}\;,
\\
 T_n  &= & \frac{p!\tau^{p+1-n}\ln{(r-\tau)}}{(n-1)!(p+1-n)!}
\; .
\end{eqnarray}
One finds that all $T_j$ terms vanish as the transition is approached
from above or below, except the term for which $j=p+1$. If $n\le p$
this term is constant and there is no discontinuity in $ f_t^{(n)}$
across the transition, while for $n=p+1$ it leads to a discontinuous 
transition with
$
\Delta f_t^{(p+1)} 
=
2 \pi g_0  p! \sin{(p+1)\phi}
$.
Therefore the first $p$ derivatives are continuous across the transition
while  the $(p+1)^{\rm{th}}$ derivative is not.
In other words, 
 to generate a discontinous transition of
order $m$ under these assumptions, 
it is necessary and sufficient that $p=m-1$, i.e.,  the 
leading behaviour of the density is
\begin{equation}
 g(r) = g_0  r^{m-1} 
\; .
\label{hjjjj}
\end{equation}
 From  (\ref{fm}) and (\ref{general}), one now has
$c_m = (-1)^l 2\pi g_0 /m$, and 
the discontinuity in the 
$m^{\rm{th}}$ derivative of the free energy is related to the density of zeros
 as
\begin{equation}
 \Delta f_t^{(m)} 
 =
 (-1)^l (m-1)! 2 \pi g_0
\; .
\label{318}
\end{equation}
This recovers the well known result that the latent heat or magnetization is related to the density
of zeros at a first-order transition through $\Delta f_t^{(1)} 
 =
 2 \pi g_0
$ \cite{LY}.

We next consider an $m^{\rm{th}}$-order diverging transition where 
\begin{equation}
 f_t^{(m)}(t)
 \sim
 |t|^{-A}
\; ,
\label{dsply}
\end{equation}
for $0 < A < 1$.
If $A=0$, we are back to the discontinuous case or the case of a logarithmic 
as opposed to power-law divergence (see the discussion 
below),
while if $A >1$, it is more appropriate to consider the transition as
$(m-1)^{\rm{th}}$ order.

Considerations similar to those in \cite{AbeF,SuzukiF}
may be used to show that in order to obtain the appropriate divergence
it is necessary and sufficient that 
\begin{equation}
 g(r) = g_0 r^{m-1-A}
\; .
\label{fo}
\end{equation}
Indeed,
from the general expression (\ref{2}), the form (\ref{dsply}) is obtained
provided (with $r = t r^\prime$)
\begin{equation}
 {\rm{Re}} \int_0^R{
                      \frac{t^A g(r) dr}{(r e^{i \phi}-t)^m}
                     }
= {\rm{Re}} \int_0^{R/t} \frac{t^{A-m+1} g(tr^\prime) dr^\prime}{(r^\prime e^{i\phi}-1)^m}
\end{equation}
is independent of $t$ as $t \rightarrow 0$.
The further condition that $g(0)=0$ gives $A < m-1$. If $m=1$, this violates the 
condition that $0<A<1$, leading to the requirement that $m \ge 2$ for a diverging 
transition. On this basis, there are no diverging first-order transitions. 
This is consistent with experience.

To demonstrate sufficiency, we put (\ref{fo}) into (\ref{2})
and use the substitution $w = r\exp{(i\phi)}/|t|$, to find,
for the $n^{\rm{th}}$ derivative of the free energy,
\begin{equation}
 f_t^{(n)}(t)
 =
  g_0(n-1)! |t|^{m-n-A}
 e^{-i(m-A)\phi} 
 I_\pm
\; ,
\end{equation}
in which
\begin{equation}
I_\pm =
2 {\rm{Re}}
\int_0^{Re^{i\phi}/|t|}{
\frac{w^{m-1-A}}{(1 \pm w)^n} dw
}
\quad
{\mbox{for}} \quad t
{\rm{\raisebox{-.75ex}{ {\small \shortstack{$<$ \\ $>$}} }}}
0\;.
\end{equation}
If $n<m$, this vanishes as  $t\rightarrow 0$,  establishing the 
continuity of the $n^{\rm{th}}$ derivative there, while
if $n=m$, one finds
\begin{equation}
f_t^{(m)}(t)
=
 - 2 g_0 |t|^{-A} \Gamma (m-A) \Gamma (A)
 \times
 \left\{
 \begin{array}{ll}
        \cos{(m-A)\phi} & \mbox{if $t<0$} \\
        \cos{\left((m-A)\phi+A \pi \right)}  & \mbox{if $t>0$}\;. 
 \end{array}
\right.
\label{end}
\end{equation}
In the case of a second-order transition,
(\ref{end}) recovers a result derived in \cite{AbeF,preAbe}.
Note that (\ref{end}) provides a direct relationship between the impact angle
and the critical amplitudes on either side of the transition. 
These critical amplitudes coincide  if the impact angle is
$ \phi = (2N-A)\pi/2(m-A)$
where $N$ is any integer.
In particular, if $m$ is even 
an impact angle of $\pi/2$ results in the symmetry of 
amplitudes around the transition. This result was already observed in the 
second-order case in \cite{AbeF}. 
The implications of (\ref{end}) are that, while this symmetry may be extended to all even-order
diverging phase transitions, it does not hold for odd ones.

If $A= 0$ in (\ref{fo}),  the singular part of the
$m^{\rm{th}}$ derivative of the 
free energy  becomes
\begin{equation}
f_t^{(m)}(t)
=
 2  (m-1)! g_0 
 \times
 \left\{
 \begin{array}{ll}
        \cos{(m\phi)} \ln{|t|}  & \mbox{if $t<0$} \\
        (\cos{(m\phi)} \ln{|t|} + \pi \sin{(m\phi)} )  & \mbox{if $t>0$}\;. 
 \end{array}
\right.
\end{equation}
This recovers a result in \cite{AbeF} if $m=2$.
Moreover, the discontinuity in the $m^{\rm{th}}$ moment across the 
transition is 
consistent with (\ref{general}) and (\ref{318}).

From (\ref{hjjjj}) and (\ref{fo}), the integrated density of Fisher zeros
is
$
 G(r) \sim r^{m-A}
$
(where $A=0$ in the case of a discontinuous transition).
For a finite system of linear extent $L$, the integrated density is defined as
$ G_L(t_j) = (2j-1)/2L^d$ \cite{JaJo04}.
Equating $G(t_j)$ to $G_L(t_j)$  leads to the scaling behaviour
\begin{equation}
 |t_j| \sim L^{-\frac{d}{m-A}}
\; .
\label{FSSF}
\end{equation}
In the diverging case where hyperscaling ($f(t) \sim \xi(t)^{d}$) holds, and
$m-A=2-\alpha = \nu d$, this recovers the usual expression,
$|t_j| \sim L^{-1/\nu}$, for finite-size scaling of Fisher zeros.
In the discontinuous case, where $A=0$, (\ref{FSSF}) yields
\begin{equation}
 \nu = \frac{m}{d}
\; .
\end{equation}
This is a generalization of the usual formal identification of $\nu$ with $1/d$, 
which applies to a first-order transition. 
Such a generalized identification
was observed at the third-order ($m=3$) discontinuous transition 
present in the spherical model in three dimensions \cite{BeKa52} as well as in the 
Ising model on planar random graphs if the Hausdorff dimension is used for $d$
\cite{fat}.

\section{Lee-Yang Zeros}
\label{sec:LY}
\setcounter{equation}{0}

In the Lee-Yang case, where there is an edge, $r_{\rm{YL}}(t)$,  to the distribution of zeros, 
the free energy is 
\begin{equation}
 f(t,h) = 2 {\rm{Re}} \int_{r_{\rm{YL}}(t)}^R{\ln{(h-h(r,t))}g(r,t) dr }
\; ,
\label{g1}
\end{equation}
where the density of zeros is written as $g(r,t)$ to display its
$t$-dependency and where the locus of zeros is 
$ h(r,t) = r \exp{(i \phi(r,t))}$.
(If the Lee-Yang circle theorem holds,  $\phi=\pi/2$ and $R=\pi$ \cite{LY}.)
The ${m^\prime}^{\rm{th}}$ field derivative of the free energy at $h=0$ is
\begin{equation}
 f^{({m^\prime})}_h(t) = 2 (-1)^{m^\prime -1} (m^\prime-1)! \frac{
                       \cos{({m^\prime}\phi)}
                      }{
                       r_{\rm{YL}}(t)^{{m^\prime}-1}
                      } 
\int_1^{\frac{R}{
                 r_{\rm{YL}}
                 }}{
                            \frac{g(xr_{\rm{YL}},t)}{x^{m^\prime}}
                         dx }
\; ,
\label{g4}
\end{equation}
having used the substitution $r=xr_{\rm{YL}}(t)$.
As in the second-order case, we assume  that $r_{\rm{YL}}(t)$ is sufficiently 
small near the transition point ($t=0$) so that the upper integral limit 
diverges
and compare with the limiting scaling behaviour in (\ref{3})
to find \cite{AbeLY,SuzukiLY}
\begin{equation}
 g(r,t) = t^{-G} r_{\rm{YL}}(t)^{{m^\prime}-1} \Phi{\left(\frac{r}{r_{\rm{YL}}(t)}\right)}
\;,
\label{g5}
\end{equation}
where $\Phi$ is an unknown function of its argument.
Similar considerations yield, for the magnetization,
\begin{equation}
 f^{(1)}_h(t,h) = 2 t^{-G} r_{\rm{YL}}(t)^{{m^\prime}-1}  
 {\rm{Re}}
\int_1^\infty{\frac{\Phi(x)}{\frac{h}{r_{\rm{YL}}(t)}-xe^{i\phi}} dx }
\; ,
\label{g6}
\end{equation}
which we may write as
\begin{equation}
 f^{(1)}_h(t,h) =  t^{-G} r_{\rm{YL}}(t)^{{m^\prime}-1}  
\Psi_\phi{\left(\frac{h}{r_{\rm{YL}}(t)}\right)}
 \; .
\label{g7}
\end{equation}
Comparison with (\ref{4}) now gives 
$
\Psi{\left(h/r_{\rm{YL}}(t)\right)}
\sim
\left(h/r_{\rm{YL}(t)}\right)^{{1}/{\delta}}
$.
The $t$-dependence must cancel in (\ref{g7}) as  $t \rightarrow 0$,
giving the small-$t$ scaling behaviour of the Yang-Lee edge
under these circumstances 
to be
\begin{equation}
 r_{\rm{YL}}(t)
 \sim
 t^{\frac{G\delta}{({m^\prime}-1)\delta -1}}
\;.
\label{3edge}
\end{equation}
When ${m^\prime}=2$ and $G=\gamma$, 
this recovers the second-order transition behaviour of (\ref{edge}).
Furthermore, (\ref{g5}) now reads
\begin{equation}
 g(r,t) = t^{\frac{G}{({m^\prime}-1)\delta -1}}  \Phi{\left(\frac{r}{r_{\rm{YL}}(t)}\right)}
\;,
\label{g50}
\end{equation}
and the expression for the magnetization in ({\ref{g7})
gives 
\begin{equation}
 f^{(1)}_h(t,h) =  t^{\frac{G}{({m^\prime}-1)\delta -1}} 
 \Psi_\phi{\left(\frac{h}{r_{\rm{YL}}(t)}\right)}
\; .
\end{equation}
Strictly, this equation of state has been derived for $t>0$, where there is an edge.
However we may assume it can be analytically 
continued into the low temperature regime, where,
taking the $h \rightarrow 0$ limit and comparing with the magnetization in (\ref{3}),
it yields 
the scaling relation
\begin{equation}
 \beta = \frac{G}{({m^\prime}-1)\delta -1}
\; .
\label{sg2}
\end{equation}
In the situation where ${m^\prime}=m$, this recovers the Griffiths-type scaling relation
(\ref{Griffiths1}),  derived in \cite{KuSa02}.

Integrating (\ref{g1}) by parts gives, for the singular part of the free energy,
\begin{equation}
 f(t,h) = 2 {\rm{Re}} \int_{r_{\rm{YL}}(t)}^R{\frac{G(r,t) dr}{he^{-i\phi}-r} } 
\; ,
\label{g11}
\end{equation}
where $G(r,t)$ is the integrated density of zeros. From 
(\ref{g5}) and (\ref{3edge}), the latter  is
$
 G(r,t)   = 
 t^{G(\delta+1)/(({m^\prime}-1)\delta-1)}
 F\left( {r}/{r_{\rm{YL}}(t)} \right)
$
in which 
$
F(x) = \int_1^x{\Phi(x^\prime)dx^\prime}
$.
Again using   $r=x r_{\rm{YL}}(t)$ in (\ref{g11}),
and taking the upper integral limit to infinity,
 one has, for the free energy,
\begin{equation}
 f(t,h)
=
t^{G\frac{\delta+1}{({m^\prime}-1)\delta-1}}
 {\cal{F}}_\phi{\left(
                         \frac{h}{r_{\rm{YL}}(t)}
                 \right)
                }
\;,
\label{ppop}
\end{equation}
where
$
{\cal{F}}_\phi{\left(
                         w
                 \right)
                }
 =
2 {\rm{Re}}
\int_1^\infty{
 {F(x) }/{(w e^{-i\phi}-x)}dx
 }
$.
The $m^{\rm{th}}$ temperature derivative of the zero-field
free energy is therefore of the form
$
 f_t^{(m)}(t)
 \sim
 t^{G(\delta+1)/(({m^\prime}-1)\delta-1)-m}
$. Comparison with (\ref{3}) then yields the scaling relation
\begin{equation}
 A = m - G\frac{\delta+1}{({m^\prime}-1)\delta-1}
\;.
\label{sg1}
\end{equation}
Together, (\ref{sg2}) and (\ref{sg1}) recover all four scaling relations
derived in \cite{KuSa02} in the more restrictive case where ${m^\prime}=m$. 
In the second-order case ($m=2$), they recover
the standard Rushbrooke and Griffiths scaling laws of (\ref{RG2}).

In fact, these laws also hold in the present case, albeit with negative 
$\alpha$ (and possibly $\gamma$).
To see this, let $f_t^{(n)}(t) \sim t^{-\alpha_n}$ and
$f_h^{(n)}(t) \sim t^{-\gamma_n}$ (so that $\alpha_2=\alpha$
and $\gamma_2=\gamma$). Since $f_t^{(m)}(t) \sim t^{-A}$, one
has, directly, that 
$
 n-\alpha_n=m-A
$.
Differentiating (\ref{ppop}) with respect to field, now gives 
\begin{equation}
 f_h^{(n)}(t)
 \sim
t^{\beta - (n-1) \beta \delta }  
=
t^{n\beta - (n-1)(m-A)}  
\;,
\end{equation}
having used (\ref{sg2}) and (\ref{sg1}) and set $h=0$. 
Now, one has
\begin{equation}
 \gamma_n= (n-1) \beta \delta -\beta
,
\quad
 (n-1)\alpha_n + n \beta + \gamma_n = n(n-1) 
,
\label{lr}
\end{equation}
which recover (\ref{RG2}) when $n=2$.%
\footnote{
It is interesting to note the restrictions imposed on $\delta$ at a higher-order transition
coming from the first equation of (\ref{lr}). For $n<{m^\prime}$, $\gamma_n$ should be 
negative, so, if $\beta$ is positive,
the best bound on $\delta$ is
$\delta < 1/({m^\prime}-2)$.
Also, the second formula in (\ref{lr}) gives, for 
$2\le n \le m-1$
and hence $\alpha_n < 0$, 
$\delta > (m-1)/\beta-1$ or $\beta > (m-1)({m^\prime}-2)/({m^\prime}-1)$.
These are no restraints in the familiar second-order case (where $m= m^\prime = 2$ and
large $\delta$ is common),
but are severe constraints at higher order.}

The formul{\ae} (\ref{3}) describe the behaviour of various moments 
as the critical point is approached tangential to the transition line (i.e., along
$h=0$). One may also be interested in the orthogonal behaviour, namely,
the $h$-dependence at $t=0$. In the case of the $h$-derivatives of free 
energy, this comes directly from (\ref{4}). For the $t$-derivatives,
we may assume the power-law behaviour (at $t=0$),
\begin{equation}
 f_t^{(j)}(h) \sim h^{s_j}
\;.
\label{sj}
\end{equation}
In the second-order case, (\ref{sj}) gives the $h$-dependency
of the internal energy and the specific heat at $t=0$ as
$
 e(h) = f_t^{(1)}(h) \sim h^{\epsilon}
$ and
$
 C(h) = f_t^{(2)}(h) \sim h^{- \sigma}
$.
These exponents are related to 
$\delta$ and $\gamma$ through (see \cite{AbeLY} and references therein)
\begin{equation}
 \epsilon = 2 - \frac{(\delta-1)(\gamma+1)}{\delta \gamma}
\;,\;
 \sigma =  \frac{(\delta-1)(\gamma+2)}{\delta \gamma} -2
\;.
\label{epssig}
\end{equation}
Following the reasoning of \cite{AbeLY}, we may argue that because there 
should be no phase transition away from $h=0$ for any $t$, the 
free energy, $f(t,h)$ in (\ref{ppop}) must be a power series in $t$ there.
So if ${\cal{F}}_\phi(w)$ involves a term, $w^q$,
the free energy involves
$
 t^{
    -G+
    (m^\prime-q) G \delta 
    /
    ((m^\prime-1)\delta-1)
   }
    h^q
$
which must be an integral power, $N$, of $t$.
This gives $q=m^\prime-[(m^\prime-1)\delta-1](G+N)/G\delta$, or the power series
\begin{equation}
 f(t,h) =
 \sum_{N=0}^\infty{
   a_n t^N h^{m^\prime - \frac{(m^\prime-1)\delta-1}{G\delta}(G+N)}
 }
\;.
\label{ps}
\end{equation}
Differentiating appropriately, putting $t=0$ and comparing with (\ref{sj}) yields
the scaling laws
\begin{equation}
 s_j = m^\prime- \frac{(m^\prime-1)\delta-1}{G\delta}(G+j)
\;.
\label{sl}
\end{equation}
In the second-order case with $m^\prime = 2$ this recovers (\ref{epssig}) with 
$s_1=\epsilon$ and $s_2=-\sigma$.

\section{Conclusions}
\setcounter{equation}{0}
\label{sec:ccl}

Different types of higher-order phase transitions have been analysed
using the zeros of the partition function.
In the Fisher case, the impact angle
is restricted by the order and nature of the transition.
For a transition with a discontinuity in $f_t^{(m)}(t)$, 
it is unclear how the system
selects from the $m$ permissible angles. 
For a divergent transition,  the impact angle determines the relevant
amplitude ratios.
Finite-size scaling is seen to hold at higher-order transitions
and the familiar formal identification of $\nu$ with $1/d$ that 
is used at first-order transitions
extends to $\nu = m/d$ for discontinuous transitions of $m^{\rm{th}}$ order.

Lee-Yang zeros, on the other hand, are appropriate to the case where
two parameters control the system. 
Here, they have been used as a route to derive scaling relations between
associated even and odd exponents, which recover 
well known formulae in the second-order case,
including the  Rushbrooke and Griffiths laws.

One of the main points of \cite{KuDo99} is that many higher-order transitions may exist 
which have not yet been identified as such.
Indeed determination of critical exponents or latent-heat-like discontinuities
is notoriously difficult from numerical work on finite systems where there is
no true transition and signals are smoothed out.
There, amplitude ratios are often more discerning 
and here we see impact angles even more so,
at least in theory. From 
the results herein, 
it would appear that analysis of the impact angle provides a very robust 
way to recognise the order of transitions.

~ \\ 
~ \\
\noindent
{\bf{Acknowledgements:}}
This work was partially supported by the EU RTN-Network `ENRAGE': {\em Random Geometry
and Random Matrices: From Quantum Gravity to Econophysics\/} under grant
No.~MRTN-CT-2004-005616.
RK thanks Pradeep Kumar for an e-mail correspondence.

\bigskip
%

\end{document}